\documentclass[jkps,preprint,fleqn,showpacs,showkeys]{revtex4}

\usepackage{graphicx}
\usepackage{amssymb}
\usepackage{amsmath}
\usepackage{bm}
\usepackage{lineno}
\usepackage{xspace}
\usepackage{cleveref}

\newcommand{\pp}{\mbox{$pp$}\xspace}
\newcommand{\pAu}{\mbox{$p$+Au}\xspace}
\newcommand{\pPb}{\mbox{$p$+Pb}\xspace}
\newcommand{\pythia}{\textsc{Pythia8}\xspace}
\newcommand{\deta}{\mbox{$\Delta\eta$}\xspace}
\newcommand{\dphi}{\mbox{$\Delta\varphi$}\xspace}
\newcommand{\pt}{\mbox{$p_{\mathrm{T}}$}\xspace}

\newcommand{\Nch}{\mbox{$N_\mathrm{ch}$}\xspace}

\begin{document}
\setcounter{page}{0}
\title[]{Exploring the string shoving model in PYTHIA8 \\for collective behaviors in $pp$ collisions}
\author{Junlee \surname{Kim}}
\email{junlee.kim@cern.ch}
\author{Eun-Joo \surname{Kim}}
\affiliation{Division of Science Education, Jeonbuk National University, Jeonju 54896}
\author{SuJeong \surname{Ji}}
\author{SangHoon \surname{Lim}}
\email{shlim@pusan.ac.kr}
\affiliation{Department of Physics, Pusan National University, Busan 46241}


\begin{abstract}
Two-particle correlations have been used to study property of a produced medium called Quark-Gluon Plasma (QGP) in relativistic heavy-ion collisions, and a long-range correlation among produced particles has been considered as a strong evidence of strongly interacting QGP. 
Strikingly, clear long-range correlations have been also observed in small collision systems like $pp$ and $p$+$A$ at a wide range of collision energies. 
A model introducing a repulsive force between flux tubes called string shoving can qualitatively describe the long-range near-side correlations in high multiplicity $pp$ collisions. 
This model implemented in the \pythia Monte Carlo event generator allows us to perform detailed study.
In this work, a full data analysis to calculate associated yields and flow coefficients is performed with \pythia events, and the results are compared with experimental results.
We find that the string shoving model overestimates long-range near-side correlations particularly in low multiplicity events.
We also introduce a method to subtract such overestimated correlations in low multiplicity to have a more relevant comparison with the data.

\end{abstract}

\maketitle

\section{Introduction}
\label{sec:intro}

A strong momentum correlation of produced particles from relativistic heavy-ion collisions is thought to be originated from an initial geometric anisotropy and hydrodynamic evolution~\cite{Busza:2018rrf}.
Two-particle correlations in \dphi (azimuthal angle difference) and \deta (pseudorapidity difference) have been measured to study azimuthal correlations among produced particles, and a clear correlation has been observed in long-range ($|\deta|>2$) where correlations from resonance decay and jet fragmentation are expected to be small.
Interestingly, such long-range correlation has been also observed in small collision systems like \pp, \pAu, and \pPb collisions over a wide range of collision energy at Relativistic Heavy Ion Collider (RHIC) and the Large Hadron Collider (LHC)~\cite{Nagle:2018nvi}.
Extensive studies have been performed to quantify the collective behavior in small collision systems, and a significant elliptic flow has been measured even in \pp collisions~\cite{Aad:2015gqa,Khachatryan:2016txc}.
Many theoretical approaches have been introduced to explain the data.
One is to apply hydrodynamic evolution developed to describe the flow in heavy-ion collisions~\cite{Weller:2017tsr}, and the other is a model of initial-state correlations among gluons~\cite{Dumitru:2010iy}.
The models considering hydrodynamic behavior generally provide a better description of the data, but it is not yet conclusive.

Recently another model called string shoving model was proposed to describe the long-range correlation in high multiplicity \pp collisions~\cite{Bierlich:2016vgw}.
This model introduces a repulsive force between flux tubes so that the flux tube expands both longitudinally and transversely.
In a high multiplicity \pp event, where many partonic interactions can occur, one string may overlap with many other strings.
The repulsion among these overlapping strings right after the collision results in correlation between particles even in a large \deta range.
In the comparison with the experimental data shown in Ref.~\cite{Bierlich:2017vhg}, this model implemented in \pythia event generator~\cite{Sjostrand:2007gs} shows a qualitative agreement in long-range correlation functions.
For more detailed quantitative comparisons, we perform a full analysis of \pythia events with string shoving and calculate quantities which have been measured in experiments.
In the following sections, we describe the analysis procedure and show results using \pythia event generator with different configurations.
The comparison with experimental results and discussion will be followed.

\pagebreak
\section{Study with \pythia}
\label{sec:ana}
The string shoving model is quantitatively studied by analyzing charged hadrons from millions of non-diffractive \pythia \pp events at $\sqrt{s}=13~\mathrm{TeV}$. 
The event multiplicity is defined with charged hadrons satisfying $|\eta|<$~2.4 and $\pt>$~0.4~GeV/$c$, which is equivalent definition to the experimental results.
To calculate two-particle correlations, we follow the formula for per-trigger yields~\cite{Khachatryan:2015lva}, which is defined as, 
\begin{align}
    \frac{1}{N_{\rm{trig}}} \frac{ {\rm d^{2}} N_{\rm{pair}} }{ \rm{d}\deta \rm{d}\dphi} = B(0,0)\frac{S(\deta, \dphi)}{B(\deta, \dphi)},
\end{align}
where $N_\mathrm{trig}$ and $N_\mathrm{pair}$ are the number of trigger particles and the number of pairs, respectively.
$S (\Delta\eta, \Delta\varphi)$ and $B(\Delta\eta, \Delta\varphi)$ are per-trigger yields in same and mixed events:
\begin{align}
    S(\deta,\dphi) &= \frac{1}{N_{{\rm trig}}} \frac{ {\rm d^{2}} N_{\rm{pair}}^{\rm same} }{ \rm{d}\deta \rm{d}\dphi},\\
    B(\deta,\dphi) &= \frac{1}{N_{{\rm trig}}} \frac{ {\rm d^{2}} N_{\rm{pair}}^{\rm mixed} }{ \rm{d}\deta \rm{d}\dphi},
\end{align}
and $B(0, 0)$ is the per-trigger yield at $(\deta,\dphi)=(0,0)$, where the pair acceptance is maximum.

\begin{figure}[tbh]
\includegraphics[width=0.49\textwidth]{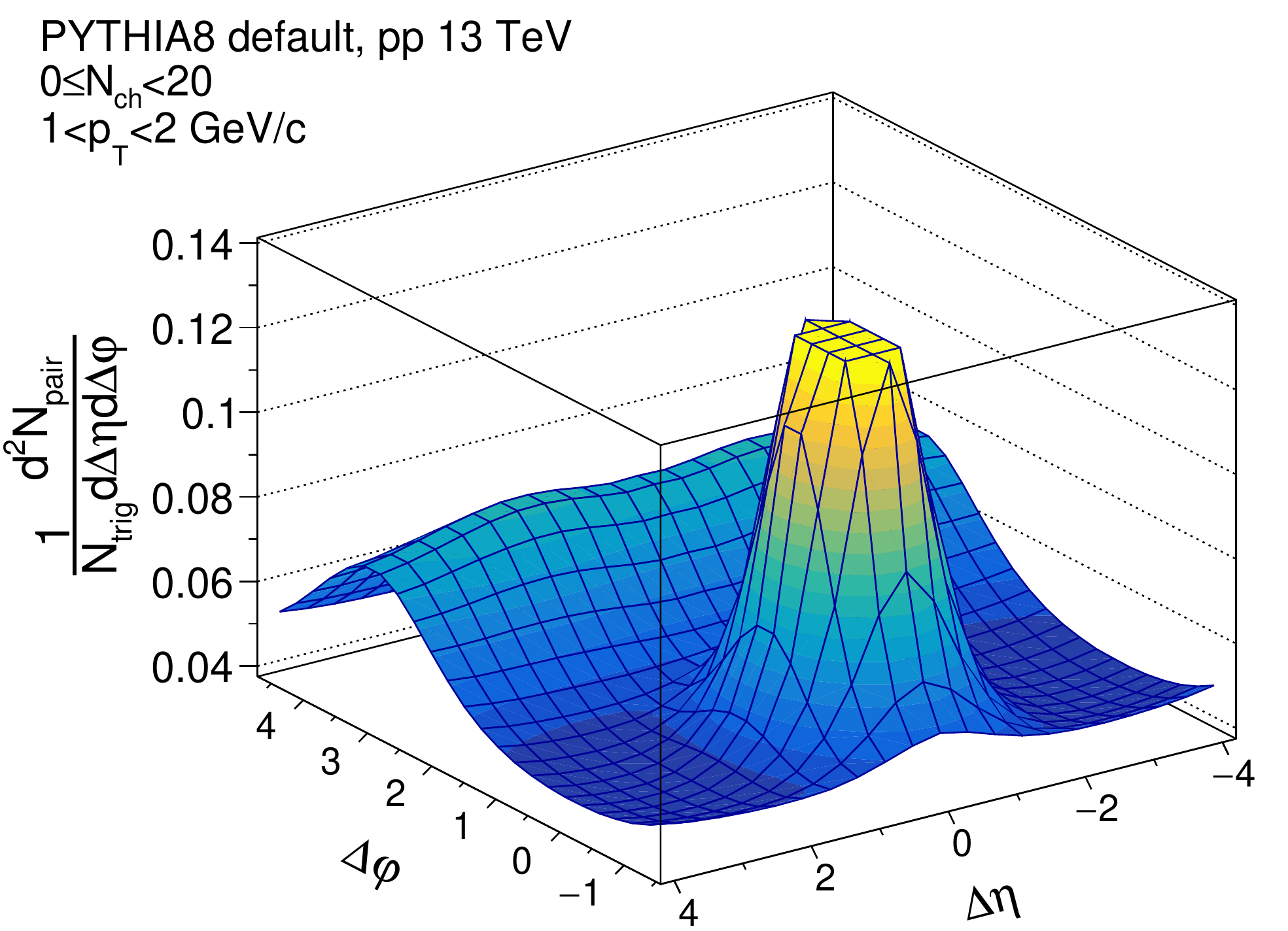}
\includegraphics[width=0.49\textwidth]{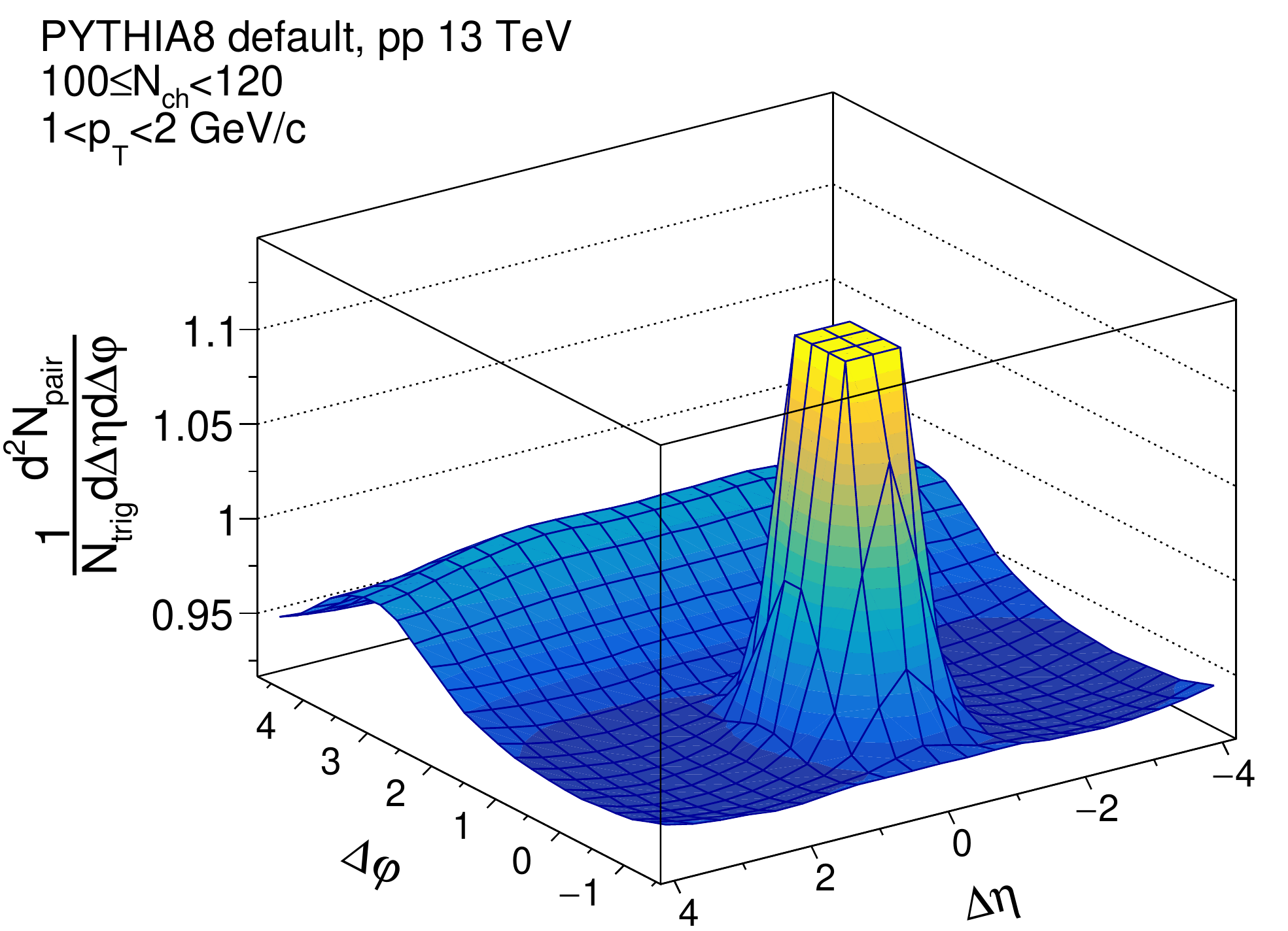}
\caption{Two-particle correlation functions for charged hadrons in $|\eta|<$~2.4 and 1~$<\pt<$~2~GeV/$c$ from low (left) and high (right) multiplicity \pythia \pp events at $\sqrt{s}=$~13~TeV with the default configuration. \Nch is calculated with charged hadrons in $|\eta|<$~2.4 and $\pt>$~0.4~GeV/$c$.}
\label{fig:2d_default_final}
\end{figure}

\begin{figure}[tbh]
\includegraphics[width=0.49\textwidth]{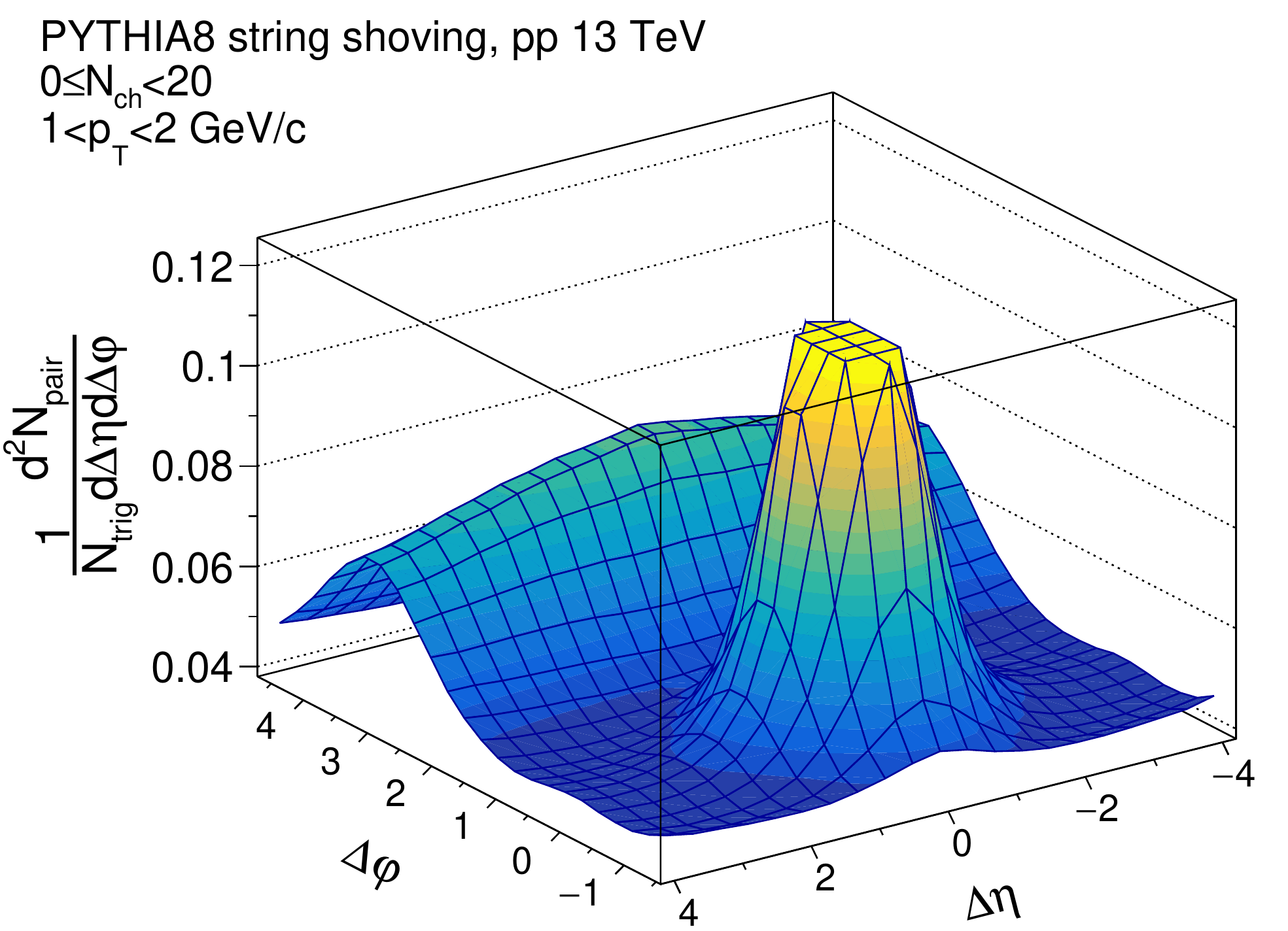}
\includegraphics[width=0.49\textwidth]{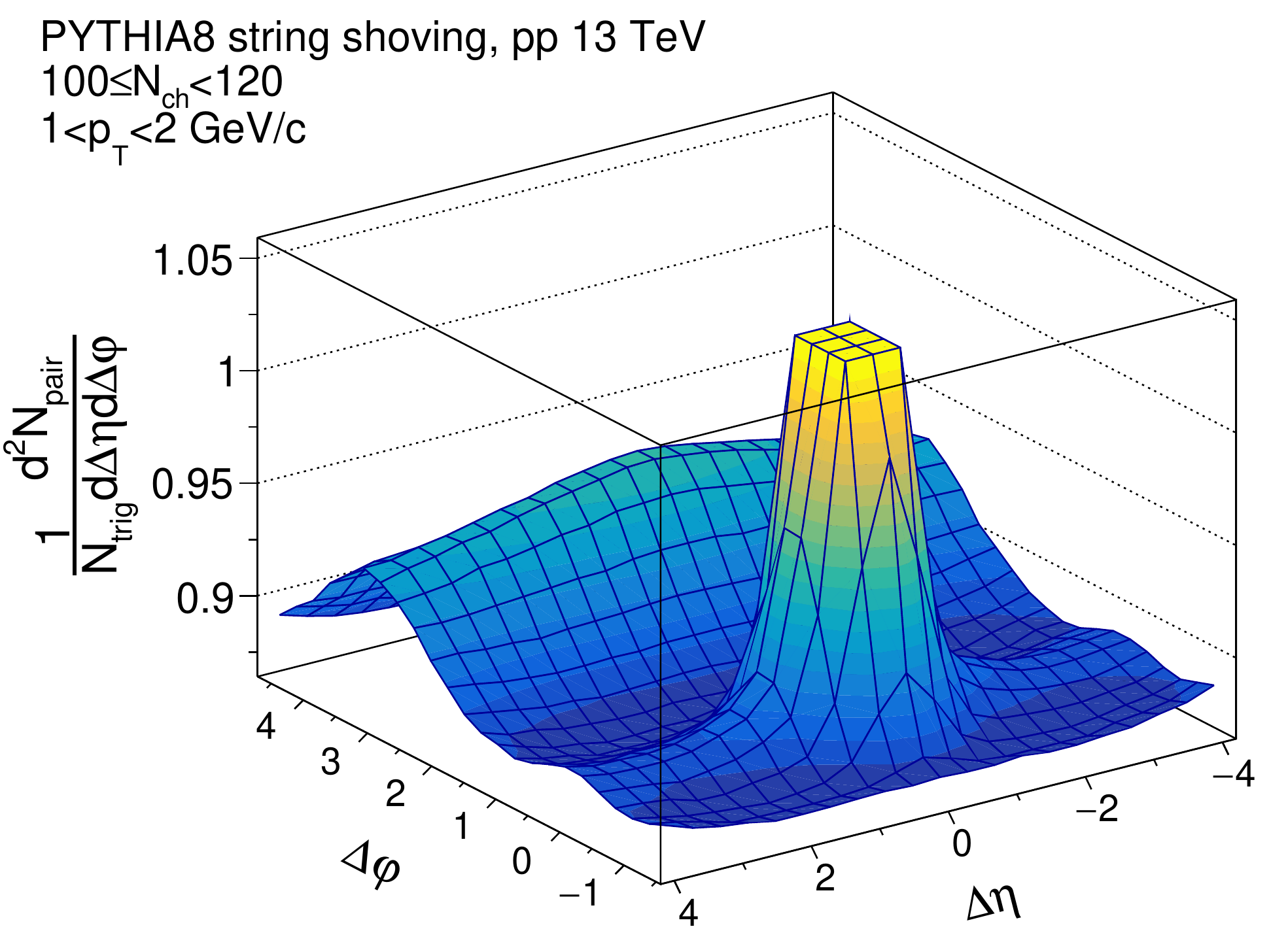}
\caption{Two-particle correlation functions for charged hadrons in $|\eta|<$~2.4 and 1~$<\pt<$~2~GeV/$c$ from low (left) and high (right) multiplicity \pythia \pp events at $\sqrt{s}=$~13~TeV with the string shoving model. \Nch is calculated with charged hadrons in $|\eta|<$~2.4 and $\pt>$~0.4~GeV/$c$.}
\label{fig:2d_shoving_final}
\end{figure}

\begin{figure}[tbh]
\includegraphics[width=0.49\textwidth]{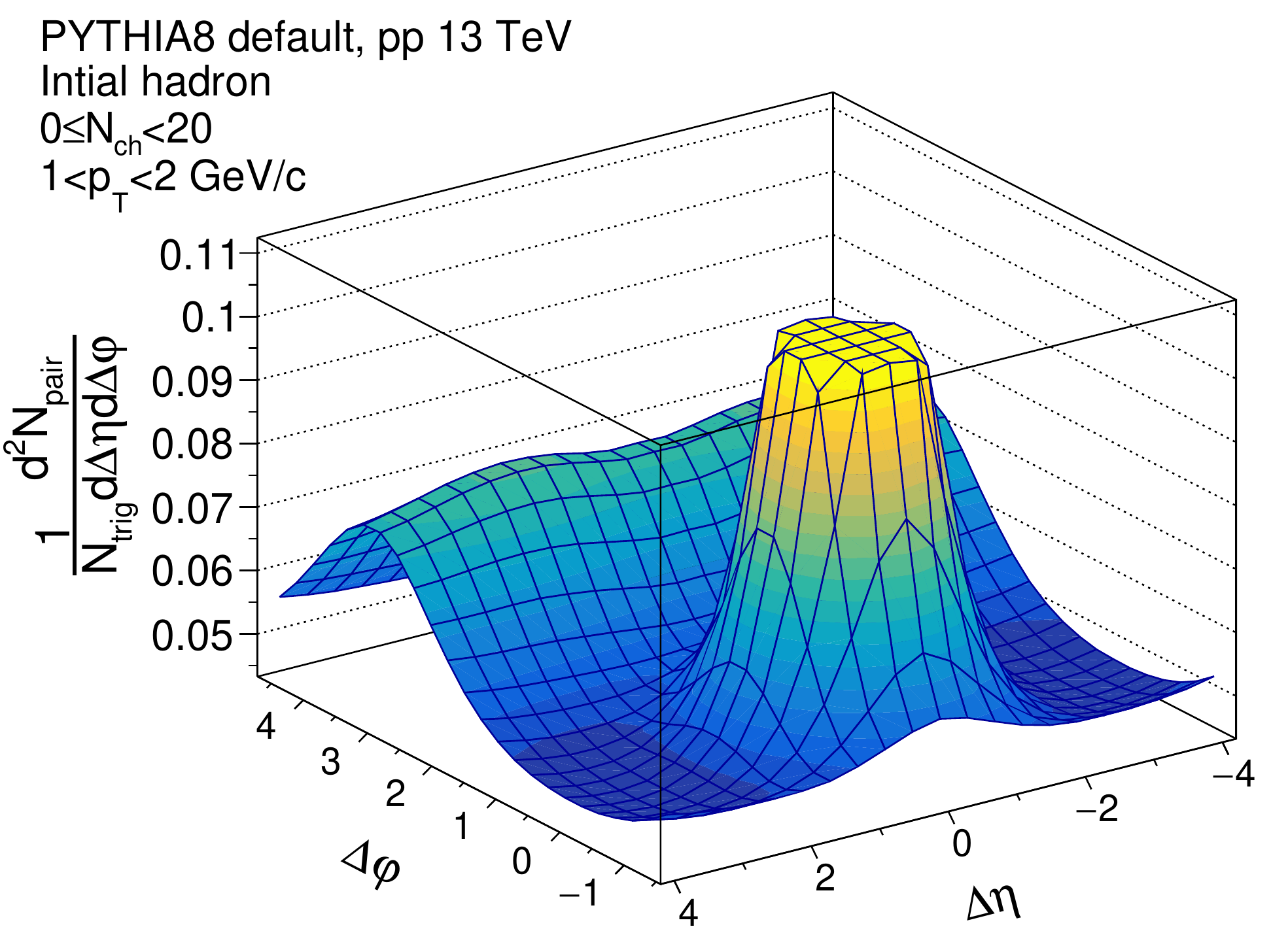}
\includegraphics[width=0.49\textwidth]{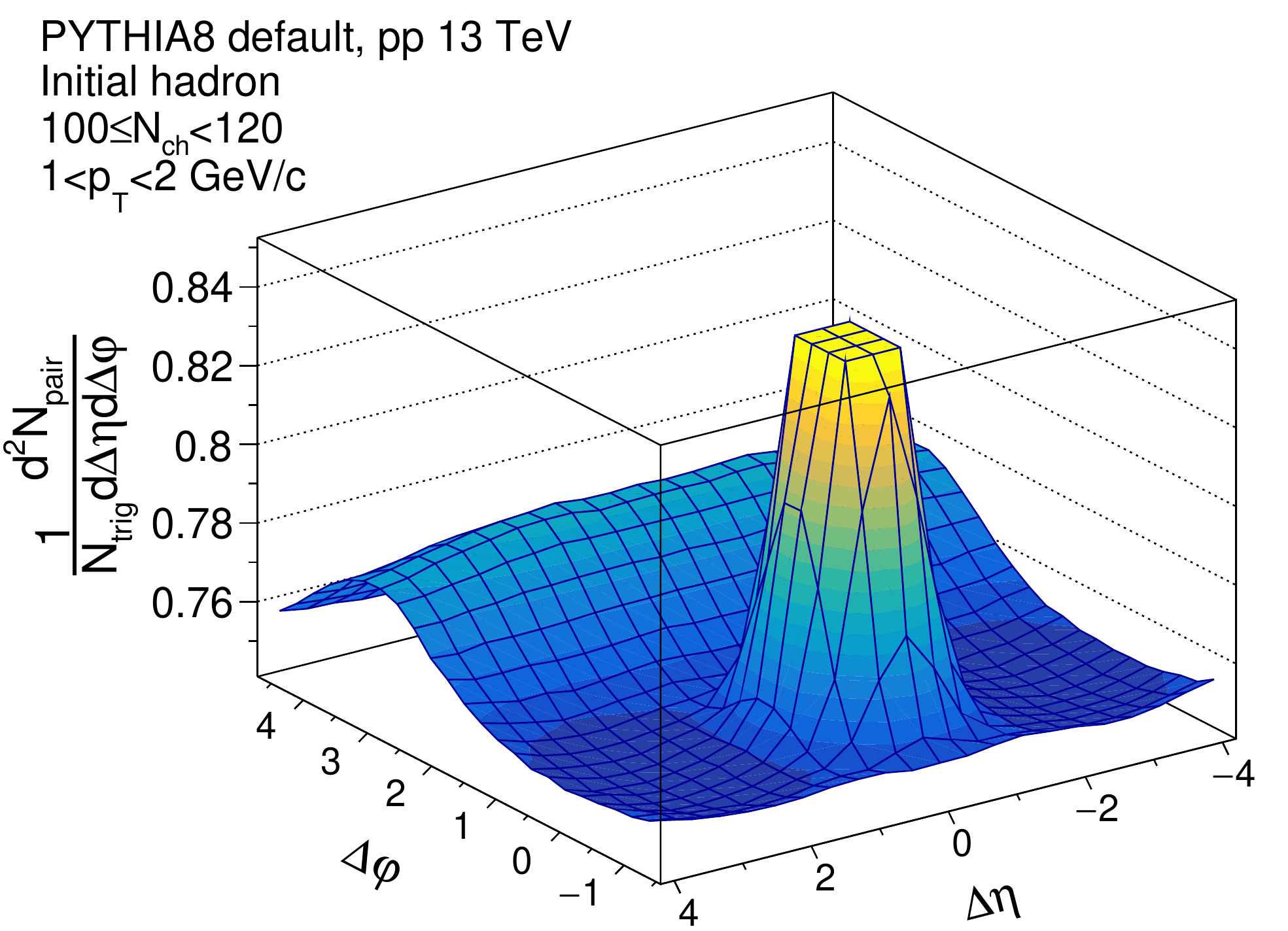}
\caption{Two-particle correlation functions for charged hadrons in $|\eta|<$~2.4 and 1~$<\pt<$~2~GeV/$c$ from low (left) and high (right) multiplicity \pythia \pp events at $\sqrt{s}=$~13~TeV with the default configuration. Initial charged particles produced directly from partons are used for the correlations, and \Nch is calculated with final charged hadrons in $|\eta|<$~2.4 and $\pt>$~0.4~GeV/$c$.}
\label{fig:2d_default_initial}
\end{figure}

\begin{figure}[tbh]
\includegraphics[width=0.49\textwidth]{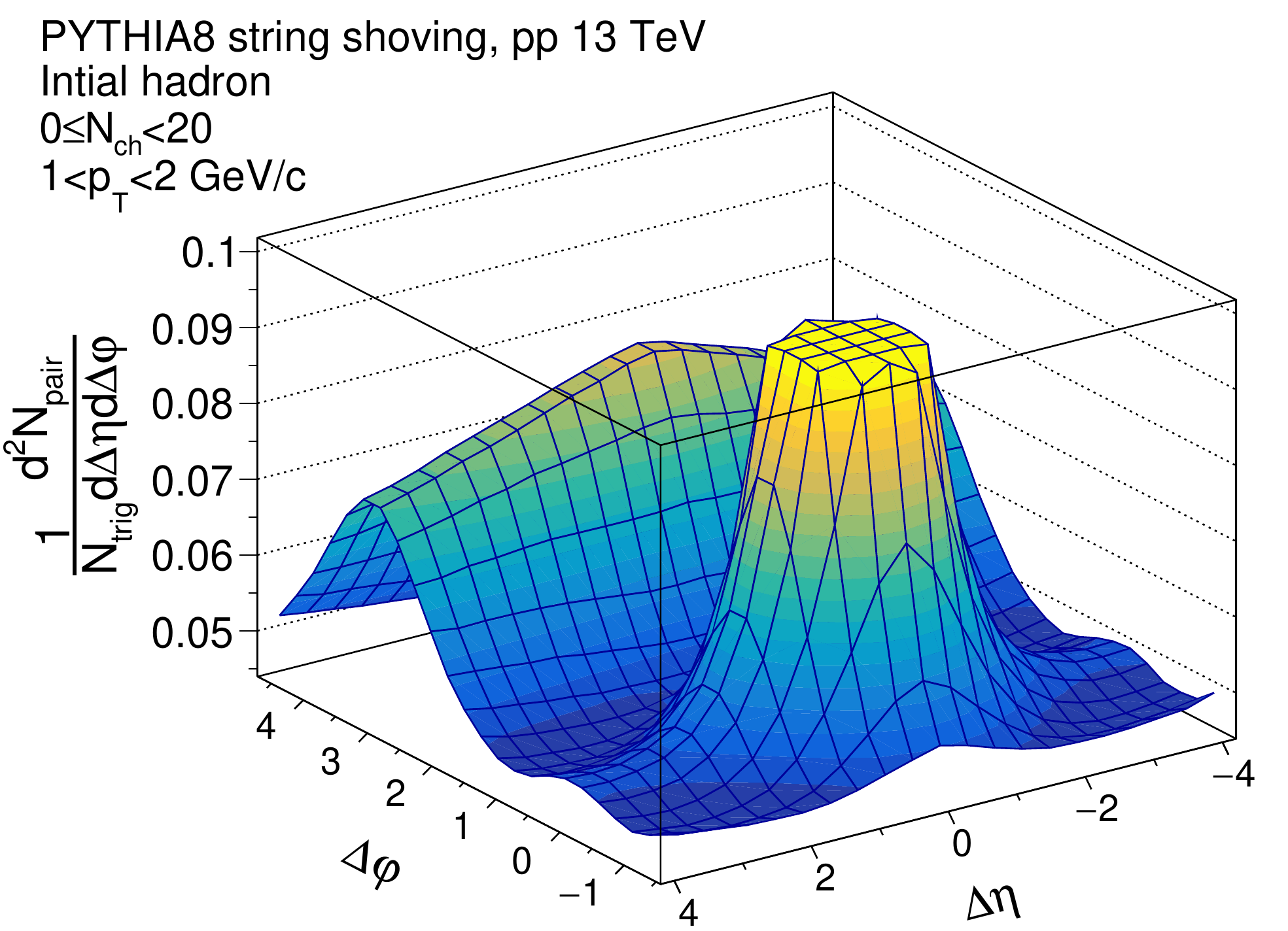}
\includegraphics[width=0.49\textwidth]{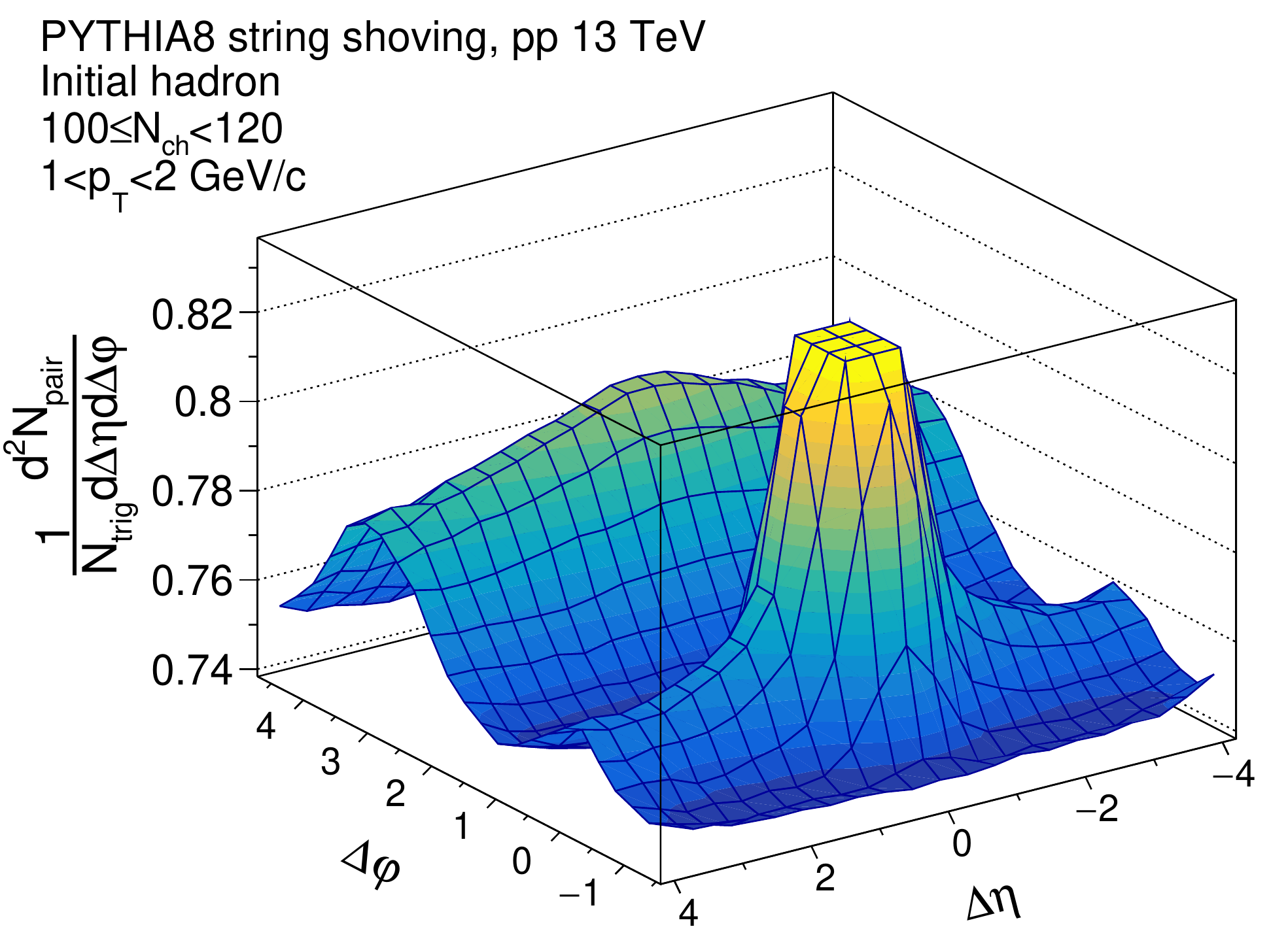}
\caption{Two-particle correlation functions for charged hadrons in $|\eta|<$~2.4 and 1~$<\pt<$~2~GeV/$c$ from low (left) and high (right) multiplicity \pythia \pp events at $\sqrt{s}=$~13~TeV with the string shoving model. Initial charged particles produced directly from partons are used for the correlations, and \Nch is calculated with final charged hadrons in $|\eta|<$~2.4 and $\pt>$~0.4~GeV/$c$.}
\label{fig:2d_shoving_initial}
\end{figure}

\Cref{fig:2d_default_final,fig:2d_shoving_final,fig:2d_default_initial,fig:2d_shoving_initial} show two-particle \deta--\dphi correlation functions with charged hadrons for 1~$<\pt<$~2~GeV/$c$ in low (left) and high (right) multiplicity \pythia \pp events at $\sqrt{s}=$~13~TeV with different configurations. 
In \pythia events with the default configuration (Figure~\ref{fig:2d_default_final}), there is no long-range near-side correlation in two different multiplicity bins as expected. 
On the other hand, The correlation functions with the string shoving option shown in Figure~\ref{fig:2d_shoving_final} clearly exhibit a long-range near-side correlation called ``ridge''.
It is also worth mentioning that a ridge structure is observed even in low multiplicity events ($0\leq \Nch <20$) with the string shoving model, which is not seen in experimental results~\cite{Khachatryan:2015lva}.
For a closer look on correlations developed in the string shoving model, correlation functions with initial charged hadrons directly produced from partons are calculated as shown in \Cref{fig:2d_default_initial,fig:2d_shoving_initial}.
Like the correlation functions with final charged hadrons, a clear long-range near-side correlation is observed in events with the string shoving, whereas no such correlation is seen in events with the default configuration.
The ridge structure with initial charged hadrons is more prominent than one with final charged hadrons, and this implies that a smearing effect in decay processes decreases the correlation strength among particles. 

One-dimensional \dphi correlation functions are constructed by averaging the two-dimensional correlation functions at a certain \deta range:
\begin{align}
\frac{1}{N_{\mathrm{trig}}} \frac{ \mathrm{d} N_{\mathrm{pair}} }{ \mathrm{d}\Delta\varphi } = \int \mathrm{d} \Delta \eta \frac{1}{\it{N}_{\mathrm{trig}}} \frac{ \mathrm{d}^{2} N_{\mathrm{pair}} }{ \mathrm{d}\Delta\eta d\Delta\varphi}.
\end{align}
For long-range correlations, the \deta range is determined as 2~$<|\deta|<$~4, where non-flow effects from di-jets and particle decay are expected to be not significant. 
The correlated part is estimated by implementing the Zero-Yield-At-Minimum (ZYAM) procedure~\cite{Ajitanand:2005jj}.
The minimum yield $(C_{\rm{ZYAM}})$ at the $\Delta\varphi$=$\Delta\varphi_{\rm{min}}$ for each \dphi correlation function is obtained by fitting the \dphi correlation function with a Fourier series up to the third term and is subtracted from the \dphi correlation function.
This procedure results in the \dphi correlation function at $\Delta\varphi_{\rm{min}}$ to be zero.

\begin{figure}[!h]
\includegraphics[width=1.0\textwidth]{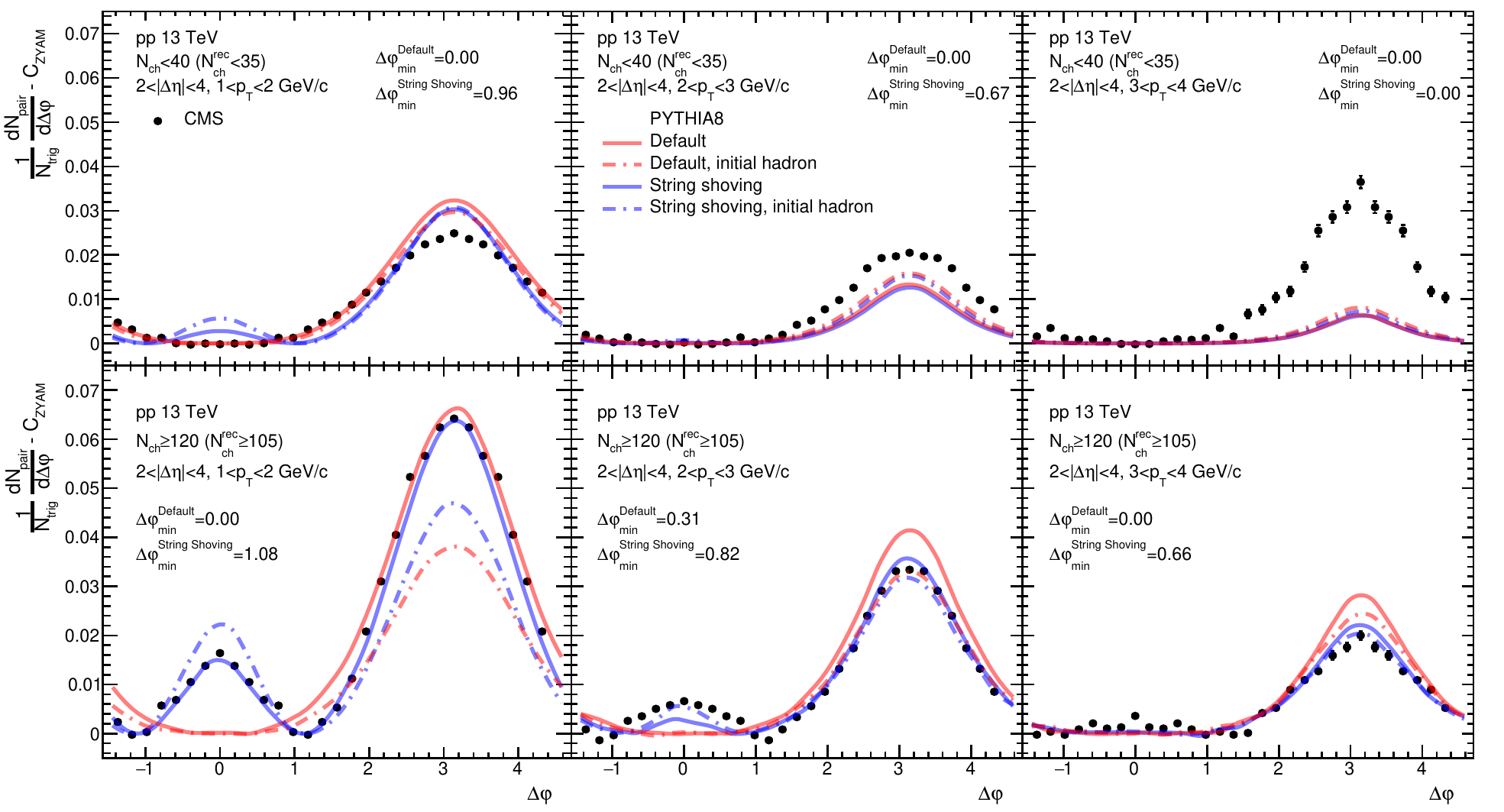}
\caption{One-dimensional long-range (2~$<|\Delta\eta|<$~4) $\Delta\varphi$ correlations in low and high multiplicity events for different \pt bins. The CMS results (closed circle)~\cite{Khachatryan:2015lva} are compared with correlation functions using initial (dashed) and final (solid) charged hadrons from \pythia events with (blue) and without (red) the string shoving. $\Delta\varphi_{\rm{min}}$ values for correlation functions with final charged hadrons are indicated.}
\label{fig:dphi}
\end{figure}

Figure~\ref{fig:dphi} shows one-dimensional \dphi correlation functions at long-range (2~$<|\deta|<$~4) after applying the ZYAM procedure for different \pt ranges in low (upper) and high (lower) multiplicity events.
\pythia events with the default configuration show no near-side structure in different \pt and multiplicity ranges. 
\pythia events with the string shoving option, on the other hand, clearly show near-side peaks.
The CMS results~\cite{Khachatryan:2015lva} are compared with the correlation functions from \pythia events, and the multiplicity ranges for \pythia events are selected $\sim15\%$ higher than the multiplicity ranges for CMS results to consider a track reconstruction efficiency.
The string shoving model overestimates the near-side correlation for 1~$<\pt<$~2~GeV/$c$ in low multiplicity events.
In the high multiplicity range, the correlation function of final charged hadrons for 1~$<\pt<$~2~GeV/$c$ in \pythia events with the string shoving nicely agrees with the CMS result.
However, the model underestimates the near-side correlation for higher \pt ranges.

\pagebreak

\section{Results} 
\label{sec:results}

\begin{figure}[!h]
\includegraphics[width=0.49\textwidth]{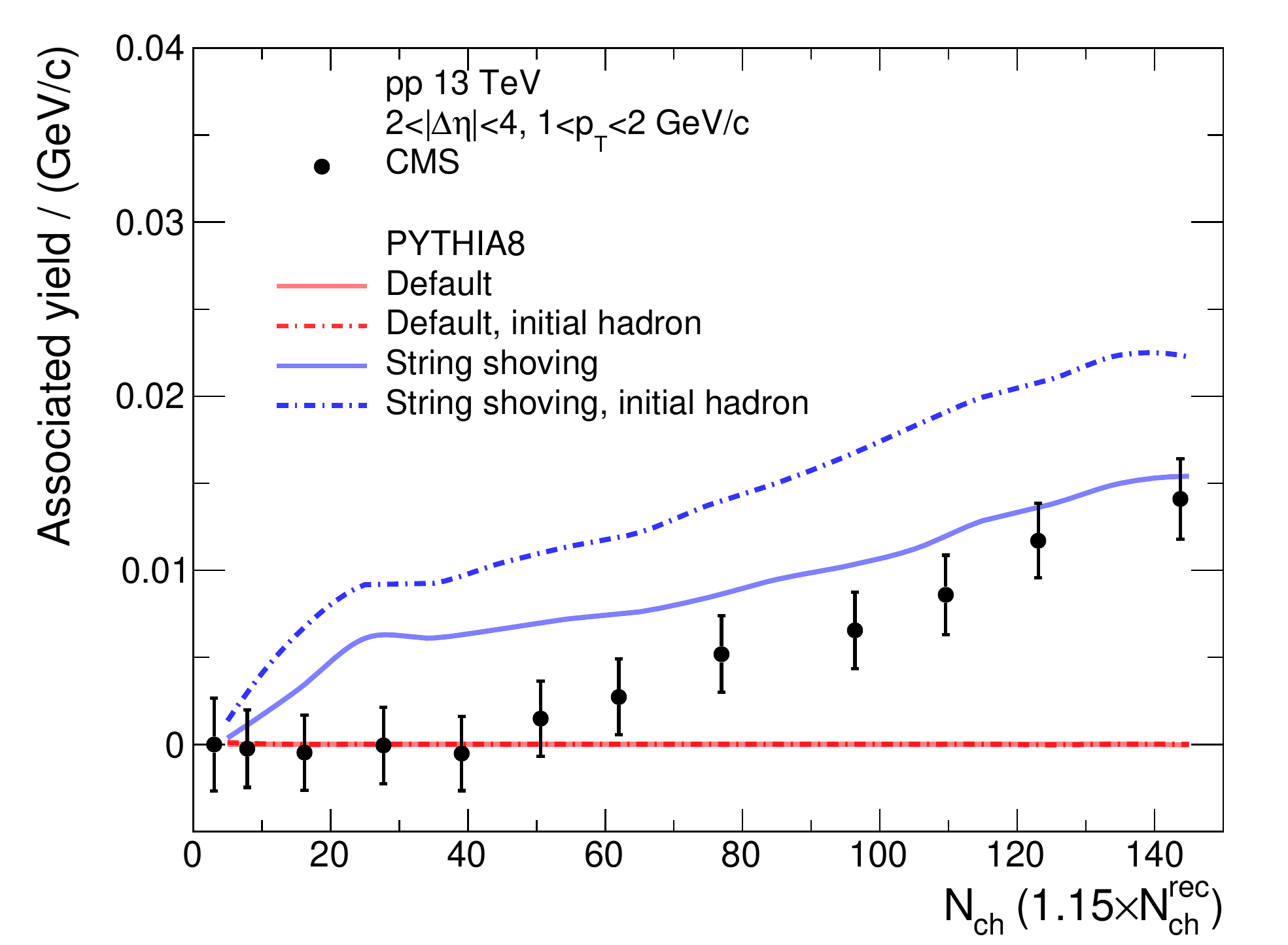}
\includegraphics[width=0.49\textwidth]{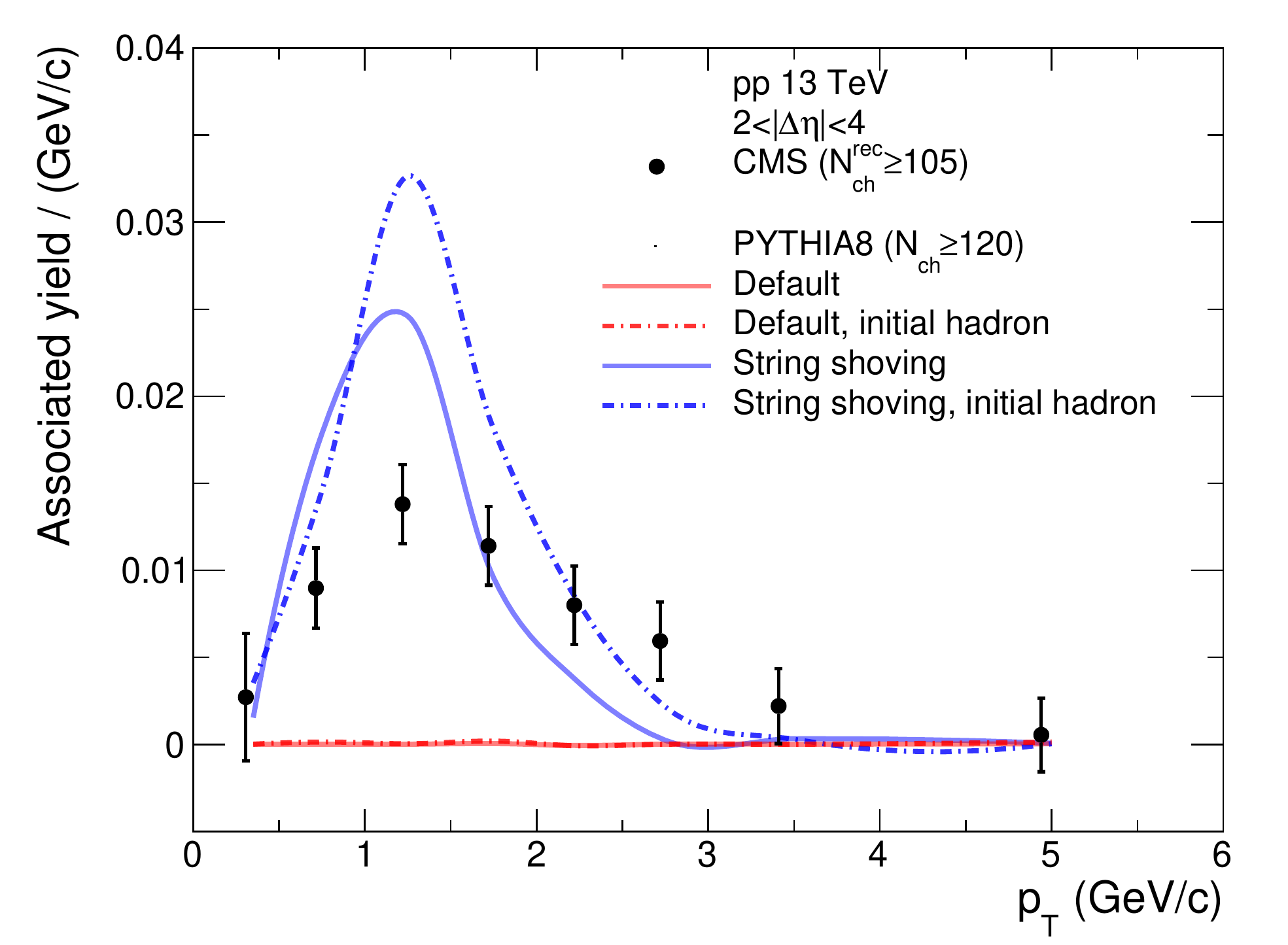}
\caption{Associated yields of long-range near-side correlation functions as a function of \Nch for charged hadrons in $1<\pt<2~\mathrm{GeV}/c$ (left) and as a function of \pt for high multiplicity events (right). The results from \pythia events with the default configuration and the string shoving model are compared with CMS results~\cite{Khachatryan:2015lva}.}
\label{fig:ay}
\end{figure}

To have more quantitative comparisons, associated yields of long-range ($2<|\deta|<4$) near-side correlation functions are calculated by integrating one-dimensional correlation functions over the region $|\dphi|<|\dphi_{\rm{min}}|$:
\begin{equation}
   Y^{\rm{assoc}} = \int_{|\Delta \varphi| < |\Delta \varphi_{\mathrm{min}}| } \mathrm{d} \Delta\varphi \left( \frac{1}{\it{N}_{\rm{trig}}} \frac{ \rm{d}\it{}N_{\rm{pair}} }{ \mathrm{d}\Delta\varphi } - C_{\rm{ZYAM}} \right), 
\end{equation}
where $\dphi_\mathrm{min}$ is the \dphi value from the ZYAM method. 
The left panel of Figure~\ref{fig:ay} shows the associated yield of long-range near-side correlation functions as a function of \Nch for charged hadrons in $1<\pt<2~\mathrm{GeV}/c$, and the multiplicity \Nch is calculated with charged hadrons in $|\eta|<2.4$ and $\pt>0.4~\mathrm{GeV}/c$.
Note that associated yields with initial charged hadrons directly produced from partons are also calculated in the same \Nch bins with final charged hadrons to use same events for associated yields with initial and final hadrons.
The associated yields from \pythia events with the default configuration (red lines) are zero for the entire \Nch range. 
In case of the string shoving model (blue lines), significant amount of associated yields are observed at the entire \Nch range, and it increases with charged hadron multiplicity.
The larger associated yields of initial hadrons than those of final hadrons may indicate that the long-range correlation built from the string shoving smears out through particle decay processes.

The CMS results~\cite{Khachatryan:2015lva} are compared with the string shoving model.
Note that the CMS results are as a function of the number of reconstructed charged particles ($N_\mathrm{ch}^\mathrm{rec}$), the CMS results are shifted along the $x$-axis ($1.15\times N_\mathrm{ch}^\mathrm{rec}$) based on the reported track reconstruction efficiency.
The associated yields of final hadrons from the string shoving model agree with the CMS result at high multiplicity events ($\Nch>120$), but it overestimates associated yields at lower multiplicity events ($\Nch<50$) indicating there is a long-range correlation from jets in the string shoving model.
For a more detailed comparison, associated yields as a function of \pt in high multiplicity events are shown in the right panel of Figure~\ref{fig:ay}.
Although the associated yields for charged hadrons in $1<\pt<2~\mathrm{GeV}/c$, shown in the left panel of Figure~\ref{fig:ay}, are comparable between the string shoving model and the CMS result in high multiplicity events, they show a different \pt dependence.
The associated yields from the string shoving model is higher than the CMS result around $\pt=1~\mathrm{GeV}/c$, and it drops more rapidly for $\pt>2~\mathrm{GeV}/c$.

\begin{figure}[!h]
\includegraphics[width=0.5\textwidth]{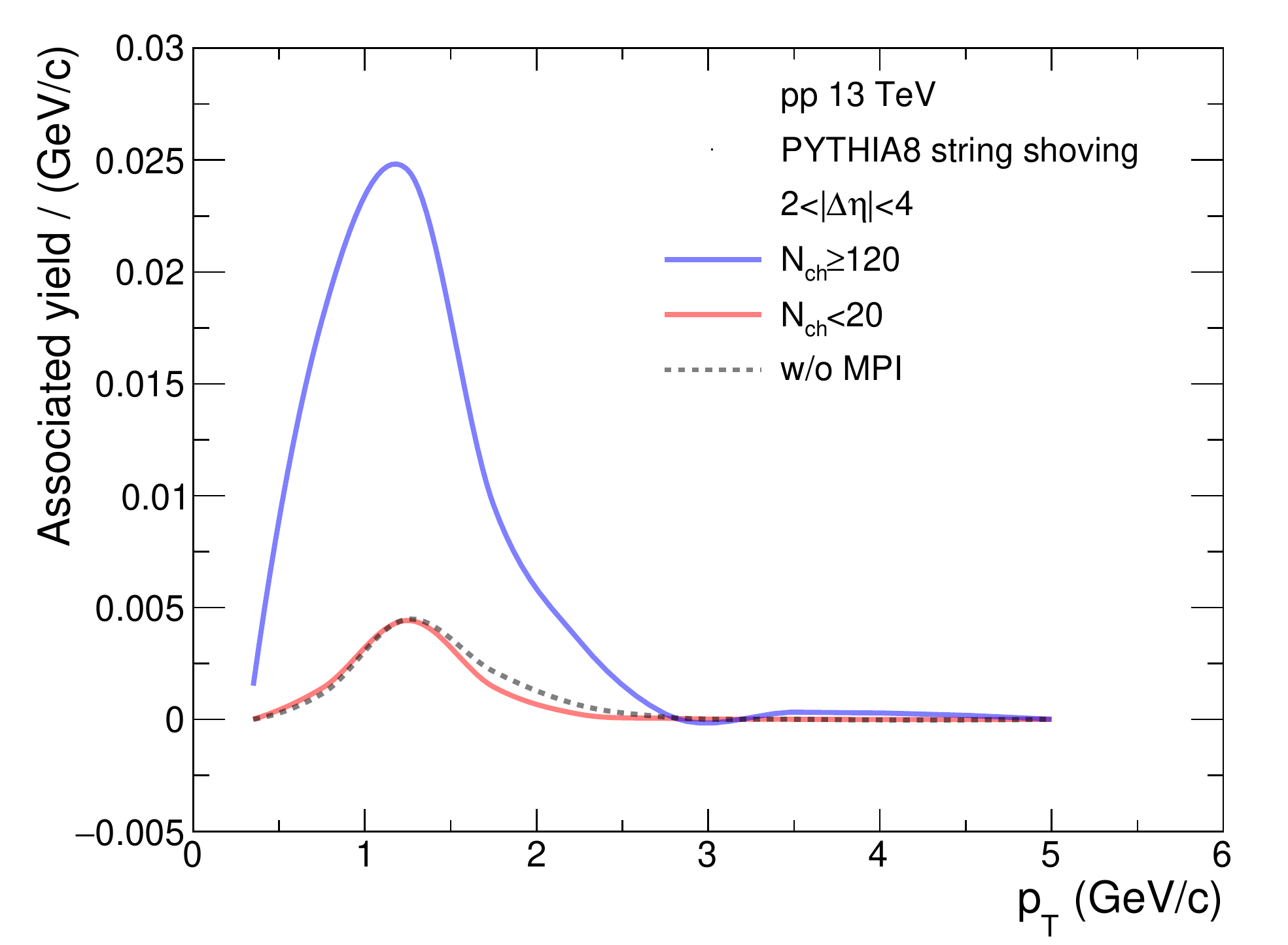}
\caption{Associated yields of long-range near-side correlations as a function of $\pt$ in \pythia with the string shoving model for low, high multiplicity events (solid lines), and events without multi-partion interaction (dashed line). }\label{fig:wompi}
\end{figure}

For further investigation on the non-zero associated yields from the string shoving model in low multiplicity events ($N_\mathrm{ch}\lesssim 50$), we generated \pythia events with the string shoving configuration but turning off the multi-parton interaction option, so charged hadrons are mostly produced from hard scattered partons.
In these events where the number of strings is expected to be small, a clear long-range near-side correlation is still observed.
Figure~\ref{fig:wompi} shows associated yields of long-range near-side correlations as a function of \pt for different event classes.
Solid lines are from normal low ($\Nch<20$) and high ($\Nch \geq 120$) multiplicity \pythia events with the string shoving configuration, and the dashed line is from inclusive events without multi-parton interaction.
The associated yields in normal low multiplicity events and events without multi-parton interaction are consistent. 
This indicates that the string shoving model introduces a long-range correlation for particles associated with jets, which is not observed in experimental results at low multiplicity events ($N_\mathrm{ch}^\mathrm{rec}<40$).

\begin{figure}[!h]
\includegraphics[width=0.49\textwidth]{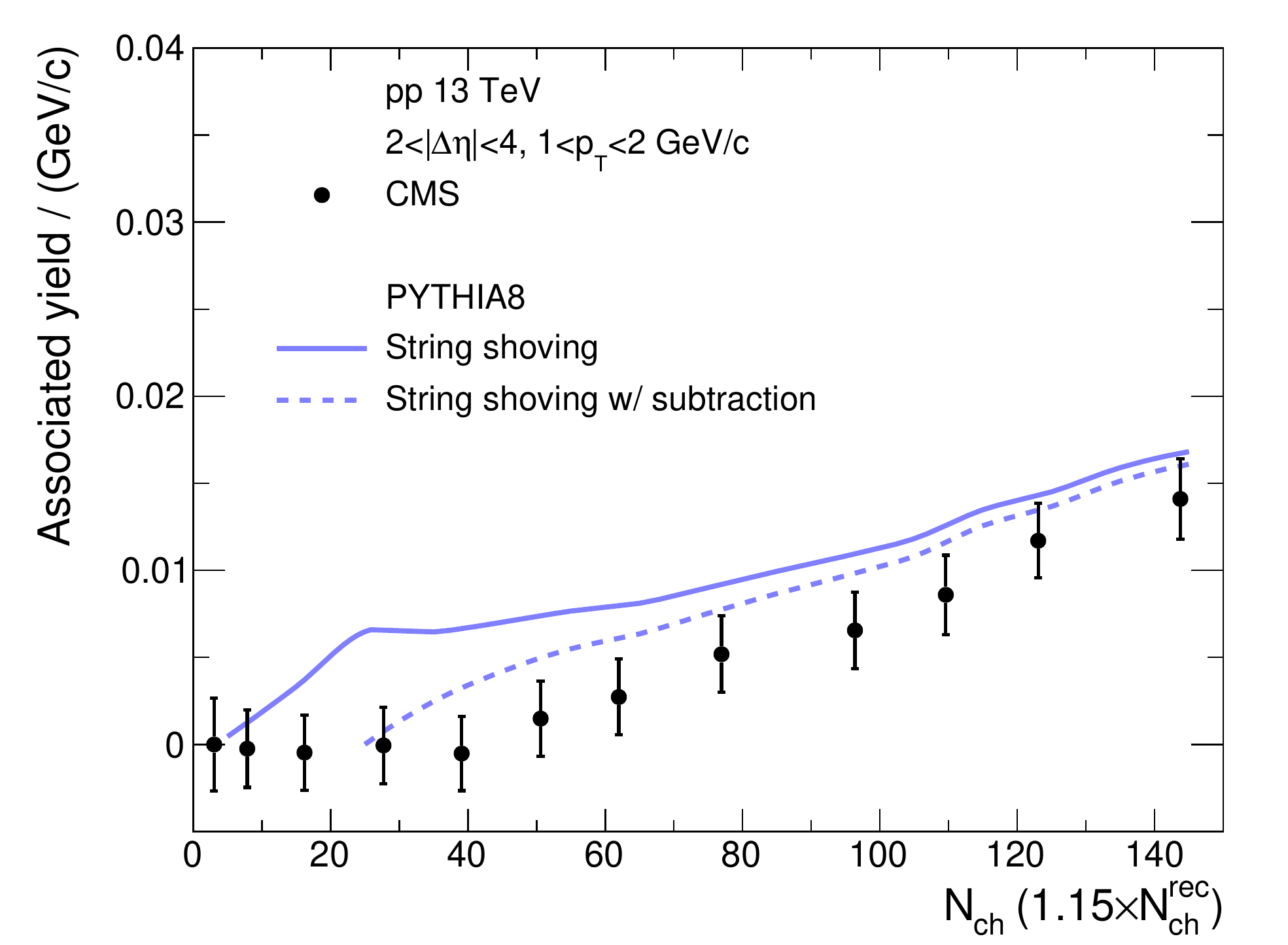}
\includegraphics[width=0.49\textwidth]{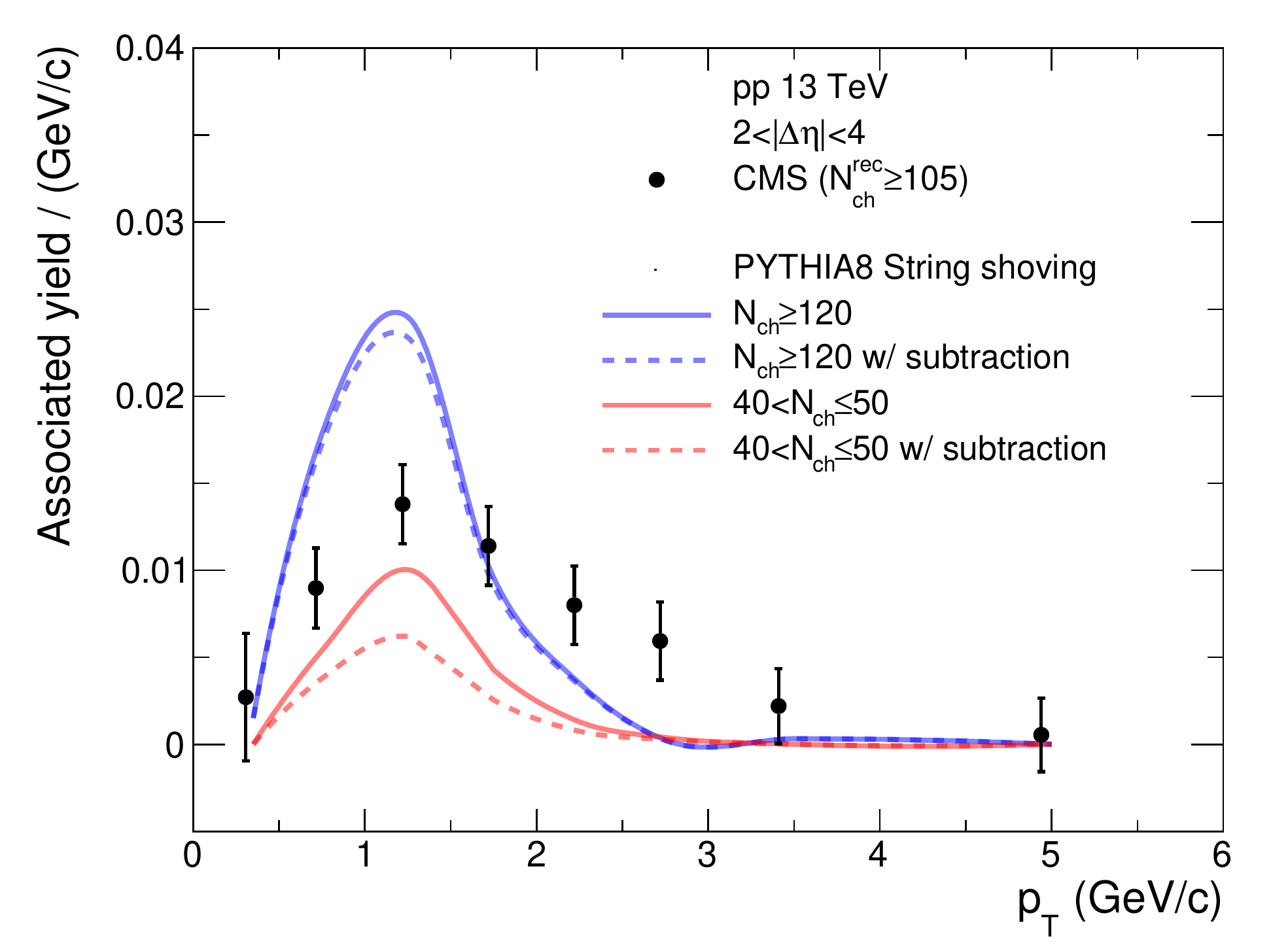}
\caption{Associated yields of long-range near-side correlation functions as a function of \Nch for charged hadrons in $1<\pt<2~\mathrm{GeV}/c$ (left) and as a function of \pt for different multiplicity bins (right). Associated yields in $20\leq\Nch< 30$ are used for the subtraction procedure.}\label{fig:aysub}
\end{figure}

We perform a further analysis to subtract associated yields from jets to have a better comparison with the data.
One simple assumption is that there are two sources of particles, jets and bulk, and there is no correlation between one particle from jets and the other particle from bulk. Another assumption is that the associated yields from jets can be estimated at low multiplicity events where jet contribution is dominant. In such case, long-range two-particle correlation functions after the ZYAM procedure can be expressed as
\begin{align}
    \frac{1}{N_\mathrm{trig}} \frac{\mathrm{d} N_\mathrm{pair}}{\mathrm{d} \dphi} = \frac{1}{N_\mathrm{trig-j} + N_\mathrm{trig-b}} \left( \frac{\mathrm{d} N_\mathrm{pair-j}}{\mathrm{d} \Delta \varphi} + \frac{\mathrm{d} N_\mathrm{pair-b}}{\mathrm{d} \Delta \varphi} \right),
\end{align}
where $N_\mathrm{trig-j}$ ($N_\mathrm{trig-b}$) is the number of trigger particles from jets (bulk), and $N_\mathrm{pair-j}$ ($N_\mathrm{pair-b}$) is the number of pairs between particles from jet (bulk).
The number of pairs from jets ($N_\mathrm{pair-j}$) in a certain high multiplicity bin can be defined as
\begin{align}
    N_\mathrm{pair-j}^\mathrm{HM} = N_\mathrm{pair-j}^\mathrm{LM} \frac{N_\mathrm{evt}^\mathrm{HM}}{N_\mathrm{evt}^\mathrm{LM}},
\end{align}
where $N_\mathrm{evt}$ is the number of events in a certain multiplicity range.
Subtracted associated yields are calculated as
\begin{equation}
   Y^\mathrm{assoc,sub} = \int_{|\Delta \varphi| < |\Delta \varphi_{\mathrm{min}}| } \mathrm{d} \Delta\varphi \left( \frac{1}{\it{N}_{\rm{trig}}^\mathrm{HM}} \frac{ \rm{d}\it{}N_{\rm{pair-b}}^\mathrm{HM} }{ \mathrm{d}\Delta\varphi } \right), 
\end{equation}
where $\it{N}_{\rm{trig}}^\mathrm{HM}$ is the number of trigger particle in high multiplicity events containing trigger particles both from jets and bulk.

Figure~\ref{fig:aysub} shows results from the subtraction method.
Associated yields in $20\leq\Nch<30$ are used for the subtraction based on the distribution of associated yields as a function of $\Nch$ from the string shoving model.
The slope of the increase of associated yields with \Nch changes in $20\leq\Nch<30$.
The jet contribution would be dominating at low multiplicity, so we assume that the associated yields in $20\leq\Nch<30$ are maximum yields from jets.
In addition, no additional multiplicity dependence is considered for yields from jets, as no strong multiplicity dependence is observed in near-side jet yields calculated using a method in Ref.~\cite{CMS:2016fnw}.

The left panel shows associated yields for charged hadrons in $1<\pt<2~\mathrm{GeV}/c$ as a function of \Nch, and the dashed line is a result from the subtraction procedure.
The subtraction procedure shows a larger impact in lower multiplicity, and the subtracted yields shows a similar multiplicity dependence with the CMS results.
In case of associated yields as a function of \pt shown in the right panel, there is no significant change in \pt dependence for the high multiplicity bin.


\begin{figure}[!h]
\includegraphics[width=0.49\textwidth]{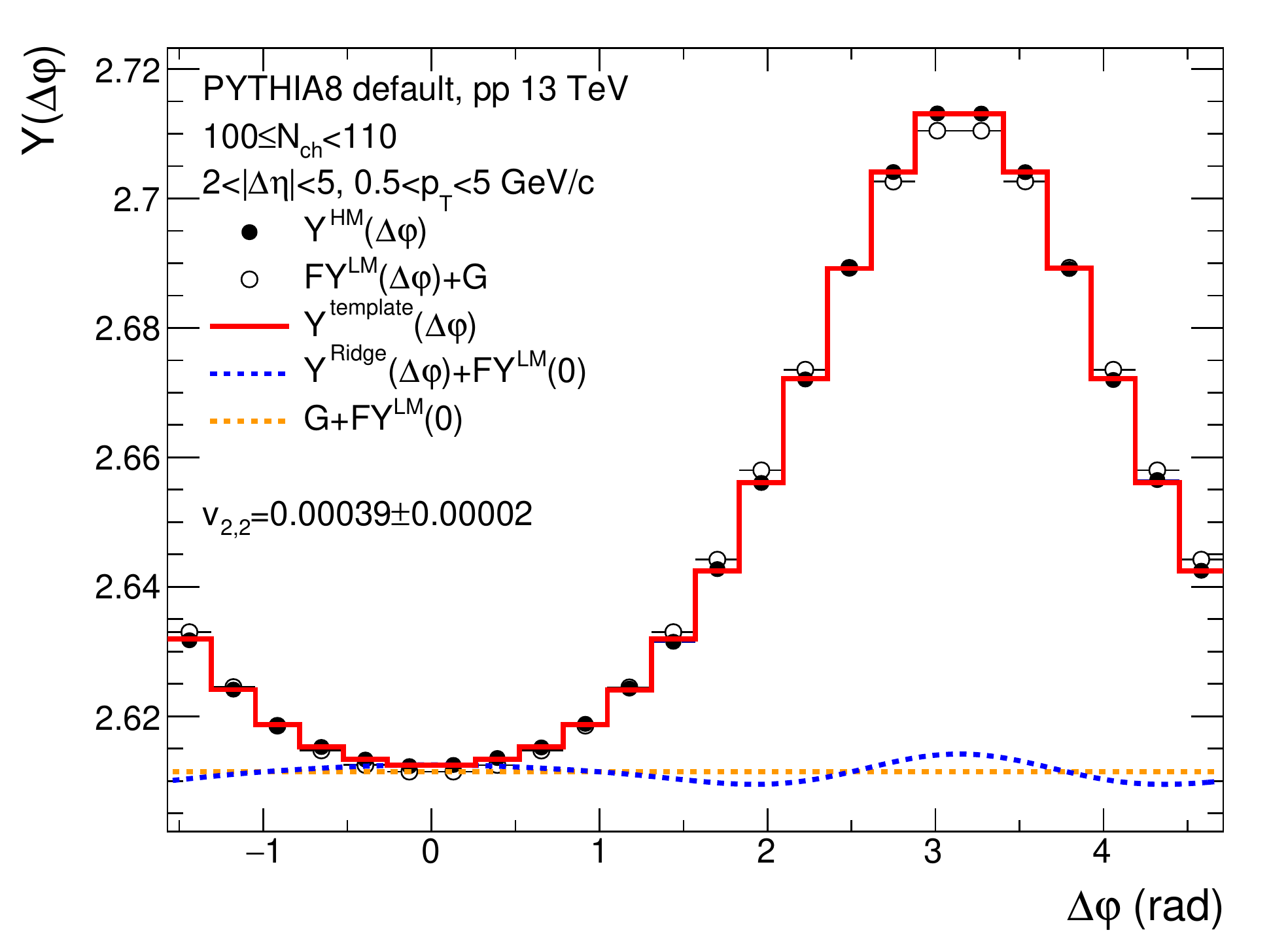}
\includegraphics[width=0.49\textwidth]{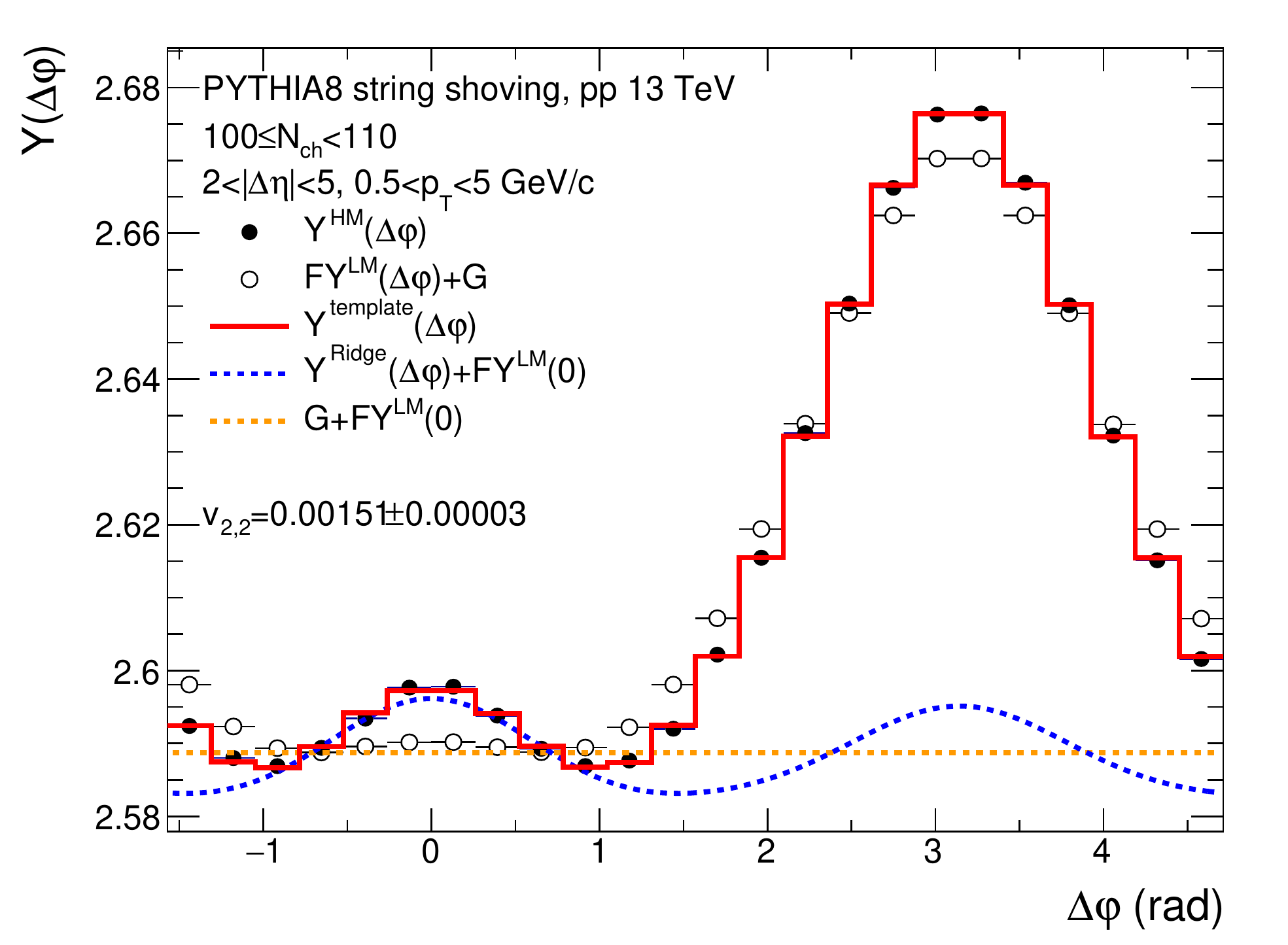}
\caption{Template fits to one-dimensional correlation functions in high multiplicity ($100\leq\Nch<110$) \pythia events with the default configuration (left) and the string shoving model (right). The correlation function in low multiplicity ($0\leq\Nch<20$) events are used for the non-flow subtraction procedure. The solid points indicate the correlation functions in high multiplicity, and the open points and curves represent different components of the template shifted along the $y$-axis by $G$ or $FY^\mathrm{LM}(0)$.}
\label{fig:templatefit}
\end{figure}

In experiments, long-range two-particle correlations are also quantified with flow coefficients.
A template fit method developed by the ATLAS collaboration~\cite{Aad:2015gqa} to subtract non-flow effects is applied to \pythia events to obtain flow coefficients.
In the template fit method, correlation functions from low multiplicity events are used to estimate non-flow contributions:
\begin{align}
    Y^\mathrm{template}(\dphi) &= Y^\mathrm{ridge}(\dphi) + FY^\mathrm{LM}(\dphi),\nonumber\\
    Y^\mathrm{ridge}(\dphi) &= G \left( 1 + \sum^{\infty}_{n=2} 2 v_{n,n} \cos(n\dphi) \right),
\end{align}
where $F$ is a scale factor for correlation functions in low multiplicity events, and $v_{n,n}$ is a coefficient related to the $n$-th order flow coefficient.
The coefficient $G$ is fixed by requiring that integrals of $Y^\mathrm{template}(\dphi)$ and $Y^\mathrm{HM}(\dphi)$ are equal. 
Based on the assumption of factorization of flow coefficients, $v_{n,n}$ from two-particle correlations between trigger particles of $p_\mathrm{T}^{a}$ and associated particles of $p_\mathrm{T}^{b}$ is $v_{n,n}(p_\mathrm{T}^{a}, p_\mathrm{T}^{b}) = v_{n}(p_\mathrm{T}^{a}, p_\mathrm{T}^{b})$.
More details on the template fit method can be found in Ref.~\cite{Aad:2015gqa}.

Figure~\ref{fig:templatefit} shows the example of the template fits to one-dimensional correlation functions of charged hadrons in $0.5<\pt<5.0~\mathrm{GeV}/c$ from high multiplicity \pythia \pp events with the default configuration (left) and the string shoving option (right).
In order to compare with the ATLAS results, slightly different selections are used for multiplicity ($|\eta|<2.5$ and $\pt>0.4~\mathrm{GeV}/c$) and long-range ($2<|\deta|<5$).
In case of the default configuration, the correlation function from high multiplicity events is well described by the scaled correlation function from low multiplicity events.
Hence, the obtained $v_{2,2}$ is very small.
With the string shoving model, a clear near-side structure is seen in the correlation function from high multiplicity events ($Y^\mathrm{HM}(\dphi)$).
However, there is also a near-side peak in the correlation function in low multiplicity events ($Y^\mathrm{LM}(\dphi)$), the scaled $Y^\mathrm{LM}(\dphi)$ take a part of the near-side correlation in $Y^\mathrm{HM}(\dphi)$.
Therefore, one can expect that the obtained $v_{n,n}$ from the template fit method will be smaller than the true $v_{n,n}$ at the same multiplicity range.

\begin{figure}[!h]
\includegraphics[width=0.49\textwidth]{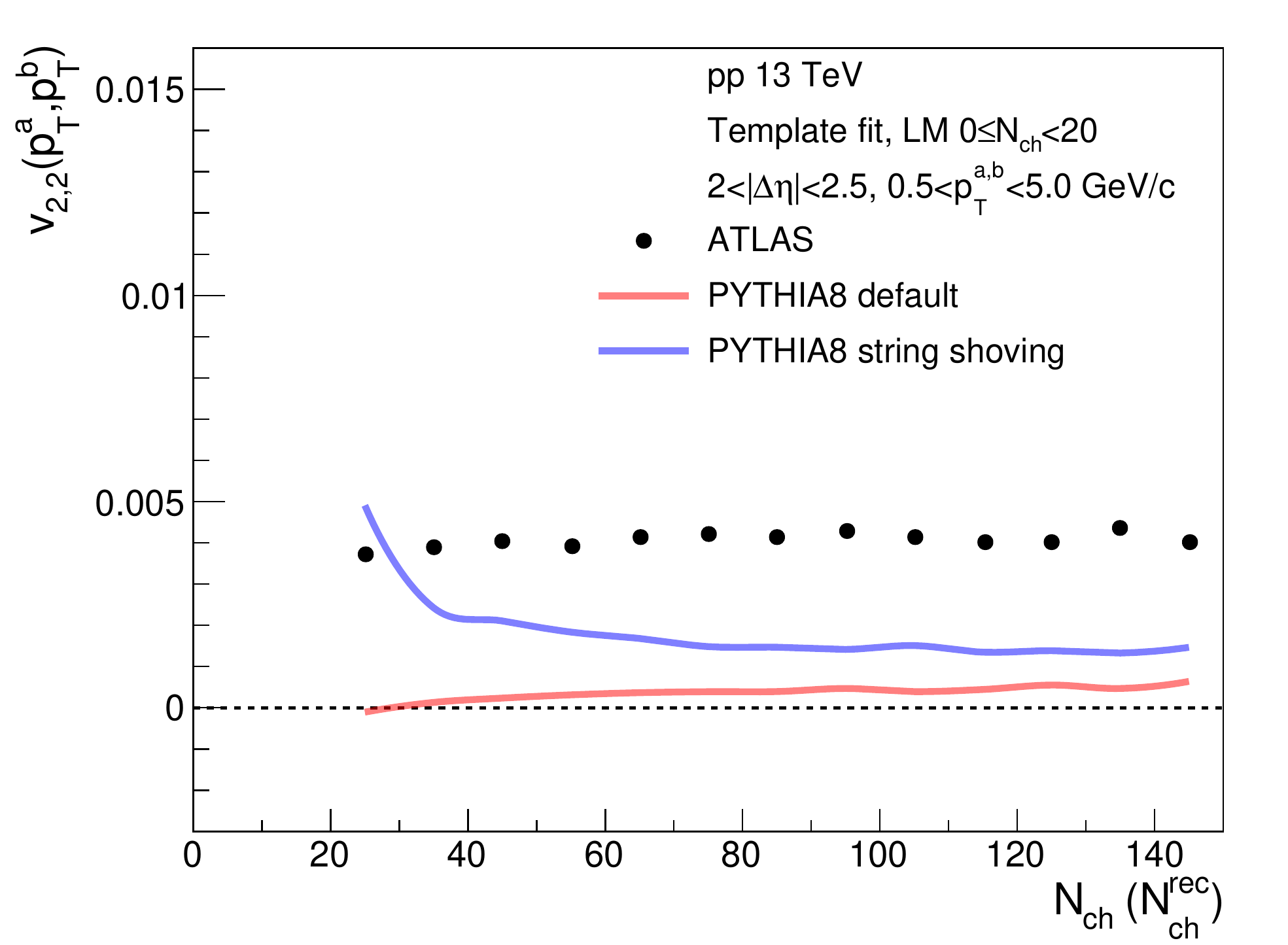}
\includegraphics[width=0.49\textwidth]{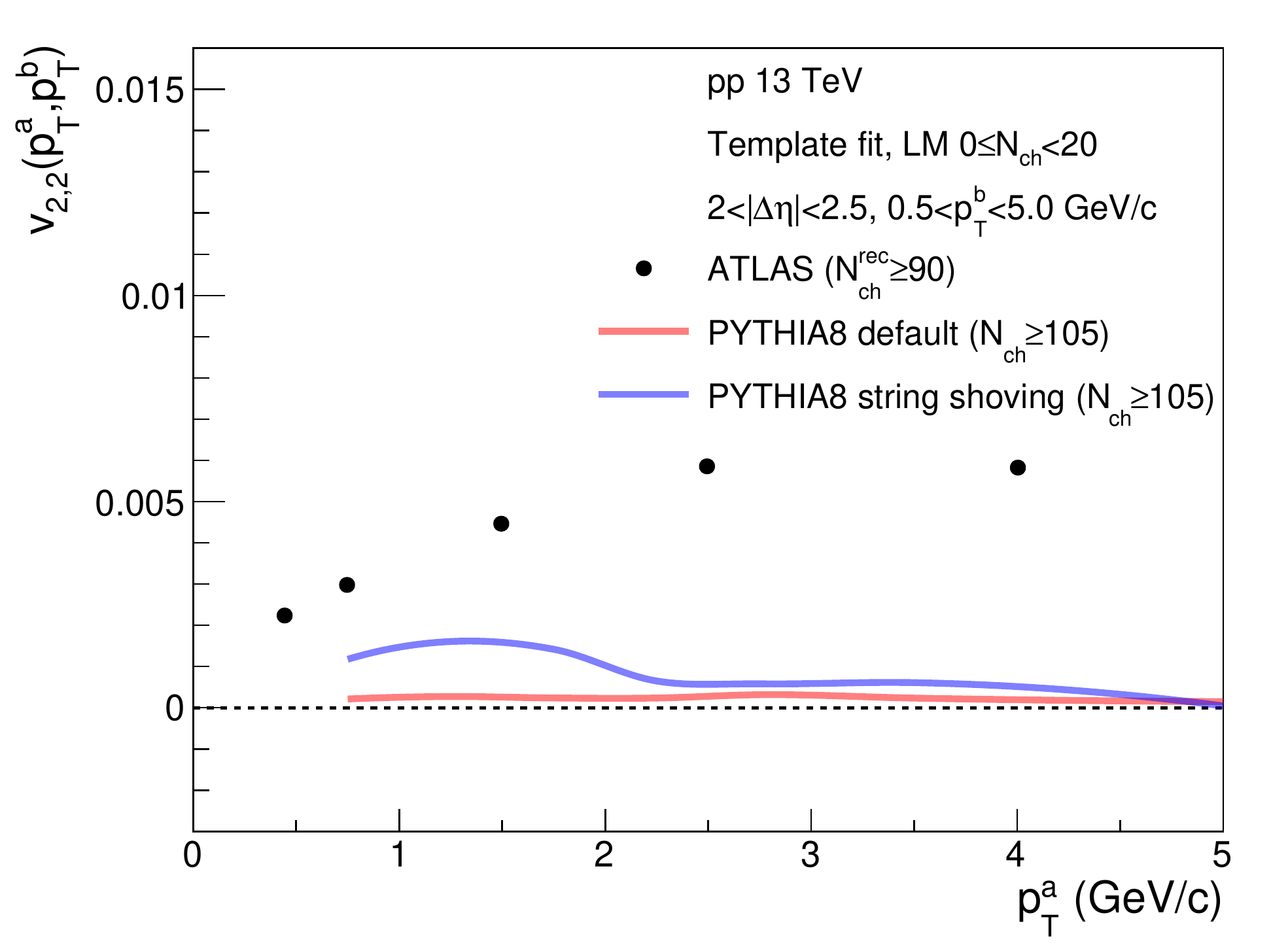}
\caption{$v_{2,2}$ obtained from the template fit method in \pythia events with the default configuration and the string shoving model, as a function of charged hadron multiplicity (left) and \pt (right). Results from \pythia events are compared with ATLAS results~\cite{Aad:2015gqa}.}
\label{fig:v22}
\end{figure}

Figure~\ref{fig:v22} shows extracted $v_{2,2}$ from the template fit method in \pythia events with the default configuration (red lines) and the string shoving model (blue lines) as a function of charged hadron multiplicity \Nch for charged hadrons in $0.5<\pt<5.0~\mathrm{GeV}/c$ (left) and as a function of \pt for high multiplicity events (right).
The $v_{2,2}$ from \pythia events with the default configuration are close to zero, and this demonstrates the non-flow subtraction method as the previous study in Ref.~\cite{Lim:2019cys}.
The results from \pythia events with the string shoving model show non-zero $v_{2,2}$ values which can be expected from long-range near-side correlation in the correlation functions.
In the comparison of the $v_{2,2}$ with ATLAS results~\cite{Aad:2015gqa}, the $v_{2,2}$ from the string shoving model is significantly lower than ATLAS results.
Based on the similar correlation functions in high multiplicity events shown in Figure~\ref{fig:dphi} and the over-subtraction of non-flow contribution shown in Figure~\ref{fig:templatefit}, the lower $v_{2,2}$ from the string shoving model is mainly due to the non-flow subtraction procedure with $Y^\mathrm{LM}(\dphi)$ showing a near-side peak.

\section{Summary}
\label{sec:summary}

In summary, a quantitative study of long-range two-particle correlations in \pythia events with the string shoving model has been performed.
Clear long-range near-side correlations are observed in the string shoving model, whereas no such structure is seen in \pythia events with the default configuration.
In the comparison with experimental data, the string shoving model qualitatively describes long-range correlations in high multiplicity events.
However, one difference is that long-range near-side correlations exist even at low multiplicity region in the string shoving model.
Due to this fact, the model overestimates associated yields of long-range near-side correlation functions, and flow coefficients ($v_{2,2}$) obtained from the template fit method are lower than
the experimental results.
A method to subtract jet correlations in associated yields is tested, and the subtracted yields of charged hadrons for $1<\pt<2~\mathrm{GeV}/c$ shows a similar multiplicity dependence with the experimental results.


\begin{acknowledgments}
S.J. Ji and S.H. Lim acknowledge support from a 2-Year Research Grant of Pusan
National University.
\end{acknowledgments}

\bibliography{paper}

\end{document}